\documentstyle[prl,aps]{revtex}
\input{psfig.sty}
\draft
\begin{document}
\twocolumn[\hsize\textwidth\columnwidth\hsize\csname@twocolumnfalse\endcsname

\title{Model of coarsening and vortex formation in vibrated granular rods}
\author{Igor S. Aranson$^1$  and Lev S. Tsimring$^2$}
\address{
$^1$ Argonne National Laboratory,
9700 South  Cass Avenue, Argonne, IL 60439 \\
$^2$ Institute for Nonlinear Science, University of California,  
San Diego, La Jolla, CA 92093-0402 }
\date{\today}

\maketitle

\begin{abstract} 
Neicu and Kudrolli observed experimentally spontaneous formation of the
long-range orientational order and large-scale vortices in  a system of
vibrated macroscopic rods.  We propose a phenomenological theory of this
phenomenon, based on a coupled system  of equations for local rods
density and tilt. The density evolution  is  described by modified
Cahn-Hilliard equation, while the tilt is described by the
Ginzburg-Landau type equation.  Our  analysis shows that, in accordance
to the Cahn-Hilliard dynamics, the islands of the ordered phase  appear
spontaneously and grow due to coarsening. The generic vortex solutions
of the Ginzburg-Landau equation for the tilt correspond to the vortical
motion of the rods around the cores which  are located near the centers
of the islands.  
\end{abstract}


\narrowtext
\vskip1pc]

Vibrated granular materials exhibit many interesting phenomena,
including formation of cellular and localized patterns, convection,
phase separation  etc.  \cite{swinney,urbach,bennaim,gollub,shinbrot,duran}.
Recently, Neicu and Kudrolli \cite{kudrolli} studied the dynamics of a
layer of long cylindrical grains (rods) subjected to vertical vibration,
and discovered a surprising phenomenon of spontaneous formation of
islands of vertically aligned rods which co-exit with the ``sea'' of
randomly packed almost horizontal rods. Subsequently, small islands
merge and form large islands which typically exhibit collective vortical
motion of rods. Near the core of the vortex this motion has a form of
solid body rotation, while farther away from the core, the angular
velocity decays. 

While the bistability and first order phase transitions leading to phase 
separation and  coarsening are typical
in granular dynamics (observed, for example, in monolayer
excitation\cite{urbach,bennaim},  vibrated powder heaping \cite{shinbrot,duran},
electrostatically-driven granular systems
 \cite{aranson}) and usually caused by
the inelasticity of grains, the emergence of the vortical motion within
the ordered phase is rather unexpected. Experiment \cite{kudrolli} shows
that rods within a vortex are tilted in the azimuthal direction, and
slowly drift in the direction of the tilt.  Authors \cite{kudrolli}
suggested that the drift occurs due to the confinement of the rod
vibration by its tilted neighbors.

In this Letter  we introduce a continuous phenomenological model of the
transition to the ordered vortical state based on the modified
Cahn-Hilliard equation governing the dynamics of local rods density and
the  Ginzburg-Landau type equation for the tilt.  Our model reproduces
qualitatively the observed  phase separation, coarsening and 
vortex formation. We 
derive the solutions for the stationary vortices and discuss their
stability.

{\it Model.} The motion of rods is described by the 
momentum conservation equation in the 
form
\begin{equation} 
\rho \left ( \frac {D {\bf v}  } {Dt} + \zeta {\bf v} \right ) 
= -\nabla p + \alpha   {\bf n}f_0(n) \rho
\label{nse} 
\end{equation} 
Here ${\bf v}=(v_x,v_y)$ is the horizontal velocity of rods, $\rho$ 
is the density, $p$ is the hydrodynamic pressure, the tilt vector 
${\bf n}=(n_x,n_y)$ is the projection of the rod  on $x-y$ plane,
and $n=|{\bf n}|$.
The term   $\alpha  {\bf n} f_0(n)\rho$, accounts for  the
driving force from the vibrating bottom exerted on the tilted rod.
According to experiments \cite{kudrolli}, the driving force
is proportional to the tilt of rods for small tilt values, but the
saturates and eventually decays to zero at $n>n_0$. 
Finally, the term $\zeta {\bf v}$
describes the momentum dissipation due to bottom friction.  Eq.
(\ref{nse}) must be augmented by the mass conservation equation in the from 
\begin{equation} 
\partial_t \rho + {\rm div}  ( {\bf v} \rho) = 0 
\label{mce}
\end{equation} 
In the following we assume that the friction is strong, and 
neglect the inertia term $\frac {D {\bf v}  } {Dt}$ with respect to 
the friction term $\zeta {\bf v}$. Thus, we can express the velocity in the form 
\begin{equation} 
{\bf v}= -\frac{1}{\zeta \rho } \left(   \nabla p 
- \alpha   {\bf n} f_0(n) \rho  \right) 
\label{vel} 
\end{equation} 
Now, substituting the velocity (\ref{vel}) into the mass conservation law 
Eq. (\ref{mce}), we obtain  
\begin{equation}
\partial_t \rho = \zeta^{-1}{\rm div}  \left( \nabla p - \alpha {\bf
n}f_0(n)  \rho\right) 
\label{mce1}
\end{equation}
To describe phenomenologically the experimentally 
observed phase separation and coarsening 
we employ the Cahn-Hilliard  approach (see for review \cite{cahn}). 
We assume that the pressure $p$ can be obtained 
from the variation of certain ``free energy'' type functional   
of the $\rho$ field
\begin{equation} 
p = \frac {\delta F}{\delta \rho}
\label{p}
\end{equation} 
We adopt the standard form of the free energy  taking into account the
local dynamics and diffusive-type coupling
\begin{equation} 
F= \int\int dx dy \left ( l^2 (\nabla \rho)^2  
+ f(\rho) \right)
\label{F}
\end{equation} 
where $l$ is the characteristic length scale related to the rod
thickness.
To account for the bistability and  phase separation,
function $f$ should have two minima separated by a maximum. 
For  simplicity we choose a quartic polynomial form of $f$, and define
$df /d\rho =  A(\rho-\rho_0)( \rho_*-\rho)(\delta-\rho)$, where 
$0 \le \rho_0 < \delta <  
\rho_*$  are characteristic rod densities depending on the driving 
acceleration \cite{minimum}. Substituting (\ref{p}) and (\ref{F}) 
into Eq. (\ref{mce1}), 
after appropriate rescaling we obtain the modified  Cahn-Hilliard equation
\begin{eqnarray}
\partial_t  \rho &=& -\nabla^2  
\left (  \nabla^2  \rho - \rho(1-\rho) (\delta-\rho) \right) \nonumber \\
&    - &\alpha 
{\rm div} \left( {\bf n} f_0(n) ( \rho+ \rho_0) \right),
\label{che}
\end{eqnarray}
where we keep the same notations for the rescaled variables and
parameters.

To close the description we need to add an equation for the 
evolution of 
tilt ${\bf n}$. For density $\rho<1$ the vertical orientation of rods
corresponding to ${\bf n}=0$ is unstable, as rods spontaneously tilt. We
assume that the growth rate of the instability 
depends on the rods packing density, so we can write
for the local dynamics
\begin{equation} 
\partial_t {\bf n}= f_1(\rho){\bf n} - |{\bf n}| ^2 {\bf n}
\label{n0} 
\end{equation} 
with $f_1(\rho)=a_0-a_1\rho$, $a_{0,1}>0$ some constants.   
As $\rho$ increases, the instability
saturates and the equilibrium tilt diminishes.
In addition, rods interact with each other, which leads to the
additional spatial derivative operator $\hat D[{\bf n}]$ in Eq.(\ref{n0}).
 Since the tilt field is not divergence-free, from the general 
symmetry considerations, in the lowest (second) order, the
``diffusion'' operator acting on ${\bf n}$, takes the form $\hat D[{\bf
n}]=f_2(\rho) \left ( \xi_1 \nabla^2 {\bf n}  + \xi_2 \nabla {\rm div} 
{\bf n}  \right)$. The coefficients $\xi_{1,2}$ in this 
expression are analogous to
the first and second viscosity in ordinary fluids (see \cite{LL}). 
Function $f_2(\rho)$ describes the decrease of the spatial coupling
strength as the rods density decreases. We assume that in the gas phase
($\rho\to 0$) the spatial coupling between the rods is small and their
tilt becomes large and uncorrelated. Accordingly, 
we set $f_2=\rho$, if $\rho>0$ and $f_2=0$ otherwise.
Finally, we include the simplest term describing coupling between the tilt
and the density gradient $\beta\nabla\rho$. Combining all these
terms  we arrive at 
\begin{eqnarray} 
\partial_t {\bf n}&=& f_1 (\rho) {\bf n} - 
|{\bf n}|^2 {\bf n} + \nonumber \\
&+&f_2 (\rho) \left ( \xi_1 
\nabla^2 {\bf n}  + \xi_2 \nabla {\rm div} {\bf n}  \right) +
\beta \nabla \rho.
\label{n1} 
\end{eqnarray} 

It is convenient to introduce new complex variable $\psi=n_x+i n_y$. Then, 
Eq. (\ref{n1})  assumes the form of the generalized Ginzburg-Landau equation
($\bar \xi=\xi_1+\xi_2/2$) 
\begin{eqnarray} 
\partial_t \psi & =&  \left (f_1(\rho)-|\psi|^2\right ) \psi +\beta (\partial_x
+i \partial_y) \rho \nonumber \\
&+&f_2(\rho) \left( \bar  \xi
\nabla^2 \psi +\frac{ \xi_2}{2} (\partial_x
+i \partial_y) ^2 \psi^* \right) 
\label{gle} 
\end{eqnarray} 

\begin{figure}[h]
\centerline{ \psfig{figure=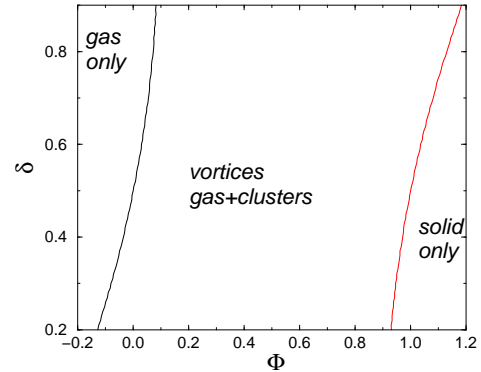,height=2.0in}}
\caption{Phase diagram in $\Phi-\delta$ plane.
Here $\Phi$ is related to the filling fraction and 
$\delta$  to the driving acceleration.
 }
\label{pdiag}
\end{figure}

{\it Phase separation.} 
Eq. (\ref{che}) exhibits phase separation only 
in a certain range of initial conditions $\rho$ and parameter $\delta$. 
The total mass conservation  yields
$\int\int \rho dxdy =S\Phi$, where $S$ is the cavity area, and 
$\Phi$ is the average density determined by the filling fraction.
Stationary solution to Eq. (\ref{che}) obeys 
\begin{equation} 
\nabla^2\rho-\rho (1-\rho) (\delta-\rho) = B,
\label{che2}
\end{equation} 
where $B=const$ is determined below. 
In the phase separation regime spatially-homogeneous 
Eq. (\ref{che2}) has three 
real roots $\rho_1 < \rho_2 < \rho_3$. 
The root $\rho_2$ corresponds to the unstable solution and $\rho_{1,3}$ to 
the stable ones. Final stage of the phase separation leads to 
one domain of the high-density phase (solid)   $\rho=\rho_3$  
of the area  $S_h$ and  the low-density one (gas)  
$\rho=\rho_1$ of the area 
$S_l=S-S_h$. From the mass conservation one obtains
\begin{equation} 
S_h \rho_3 + (S-S_h) \rho_1 = S \Phi.
\label{mce2}
\end{equation} 
Here we neglect the interfacial contributions assuming that $S_l,S_h \gg 1 $. 
In addition, Eq. (\ref{mce2}) must be augmented by the 
condition that the free energy densities of the both phases are equal, 
which is expressed by the relation (so-called area rule, see e.g. 
\cite{meerson}) 
\begin{equation} 
\int _{\rho_1}^{\rho_3} \left[\rho (1-\rho) (\delta-\rho) -B) \right] d \rho
=0.
\label{area}
\end{equation} 
Eq. (\ref{area}) and  Eq. (\ref{che2}) fix  the value 
of  $B$, and,  correspondingly the roots $\rho_{1,3}$. 
>From Eq. (\ref{mce2}) it follows that the solution with two phases is possible 
if $\rho_1 < \Phi< \rho_3$. It defines the phase separation region in 
Fig. \ref{pdiag}. 

Let us consider 
radially symmetric {\it vortex} solutions to Eq. (\ref{gle}). 
In this case  
$\rho$ is a function of the polar radius  $r$, and $\psi $ can be
expressed as 
$\psi = \exp( \pm  i \theta)  w(r)$, 
where $w$ is a complex function, and $\theta$
is the polar angle (for definiteness we take sign $+$). Using 
$\partial_x
+i \partial_y= \exp( i \theta) ( \partial_r +  i /r  \partial_ \theta)$,
 we obtain  from  Eq. (\ref{gle}) 
\begin{equation} 
\partial_t w=f_2(\bar \xi \nabla^2_r w + 
\frac{ \xi_2}{2} \nabla^2_r w^*) + f_1  w-|w|^2 w +\beta \rho_r 
\label{gle1}
\end{equation}  
where $\nabla^2_r = \partial _r^2 + r^{-1} \partial _r -r^{-2}$ is the
radial Laplacian operator.  
For $\xi_2, \beta=0$, Eq. (\ref{gle1}) possesses  a stationary solution in
the form $w=W(r) \exp( i \phi_0)$ with the real positive magnitude $W$
and an arbitrary constant phase $\phi_0$. The terms $\propto \xi_2, \beta$ 
still permit the constant phase solutions, but they destroy the
continuous phase degeneracy.  Indeed, Eq. (\ref{gle1}) for $\beta=0$ yields
\begin{eqnarray} 
 f_2 ( \bar \xi +  \frac{ \xi_2}{2}  \cos 2 \phi_0 ) 
\nabla^2_r W  + f_1W - W^3  =0  \label{gle2}  
\end{eqnarray} 
and $\sin 2 \phi_0 = 0$. 
Solutions exist only for $\phi_0 = 0, \pi $ or $\pm \pi/2$. Function $W$ 
describes the standard (and well-documented) vortex solution to the
Ginzburg-Landau equation with the property $W \to f_1^{1/2}$ for 
$r \to \infty$ 
and $W \sim r $ for $r \ll 1$  (for its rational approximation see e.g. 
\cite{pismen}).  
Solutions 
with $\phi_0=0,\pi$ describe sinks (sources)  with zero circulations, 
whereas the solutions with $\phi_0=\pm \pi/2$ are vortices with 
non-zero circulations. The sign of $\phi_0 $ determines the 
direction of rotation. Near the vortex core  $W \sim r$, 
which corresponds to the solid body rotation, as velocity 
$v_\theta \sim W $. Far away from the core the vortex exhibits
differential rotation.

For $\beta \partial_r \rho  \ne  0 $,  
Eq. (\ref{gle1})  
has constant phase solutions  only with $\phi_0=0, \pi$, i.e. with {\it zero 
circulation}.
However, it does not  guarantee the  selection of this solution in the bulk 
of large islands where the density gradient is small.  
It is easy to show that solutions with $\phi_0=0$ 
are   energetically unfavorable with respect to rotating
solutions with  $\phi_0=\pm\pi/2$ if $\beta \partial_r \rho$ is small. Indeed, 
Eq. (\ref{gle1}) for $\beta\partial_r \rho=0$ can be written in the form 
\begin{equation} 
\partial _t w= -\frac{\delta U}{ \delta w^*} 
\label{gle3} 
\end{equation} 
with the free energy functional
\begin{eqnarray} 
U &=&  \int  d{\bf r} [
 f_2 \bar \xi (|\partial_r w|^2 +r^{-2}|w|^2)  \nonumber \\ 
&+& \frac{ f_2 \xi_2}{4} \left(  ((\partial_r+r^{-1})w^*)^2 + c.c. \right) 
 -f_1|w|^2 + | w|^4   ] 
\label {U}
\end{eqnarray} 
Substituting vortex solution $w=W(r)\exp(i\phi_0)$ in Eq. (\ref{U}), 
one obtains 
after integration (since the calculation of the vortex energy is rather 
straightforward, we refer interested readers to  Ref. \cite{pismen}, p. 11). 
\begin{equation} 
U =  f_1 f_2 (\bar \xi + \xi_2/2 \cos 2 \phi_0 ) \log R/r_0 + const 
\label{U1} 
\end{equation} 
where $r_0 \sim O(1) $ has the meaning of core radius, and $R $ is  
the outer  cutoff radius of integration. As one sees from 
Eq. (\ref{U1}), for physically realizable case $\xi_2>0$, the vortices
with $\phi_0=\pm \pi/2$ have lower energy, and therefore are more 
energetically favorable, and are  selected in dynamics.
In a general case, the vortex solution would
have a radius-dependent phase $\phi_0$. Near the center where the
density is almost constant, the phase would be close to $\pm\pi/2$, and
near the island border where the density decreases rapidly, the phase
should approach $0$ or $\pi$. This scenario suggests that the azimuthal
velocity should grow with radius near the core, and decrease near
edge of the vortex, which is confirmed by our numerical simulations (see
below) and agrees with experiments.  In addition, one can show
that for $\beta>0$ the term $\beta\nabla \rho $ considered as a small
perturbation, leads to the drift of the
vortex core towards the gradient of density, and therefore it stabilizes
the vortex core near the center of the island. 

{\it Numerical simulations}   
of the  Cahn-Hilliard equation  (\ref{che})
were performed  using an FFT split-step method, and 
the Ginzburg-Landau equation (\ref{gle}) was solved using explicit method. 
The domain of integration was $100\times100$ dimensionless units with periodic
boundary conditions, 
number of mesh points/FFT harmonics was $256\times256$. 
As initial conditions we used $\rho\approx \delta$ with with small
amplitude noise  and random initial conditions for $\psi$. 
Selected results are presented in Fig. 2-5.

\begin{figure}[h]
\centerline{ \psfig{figure=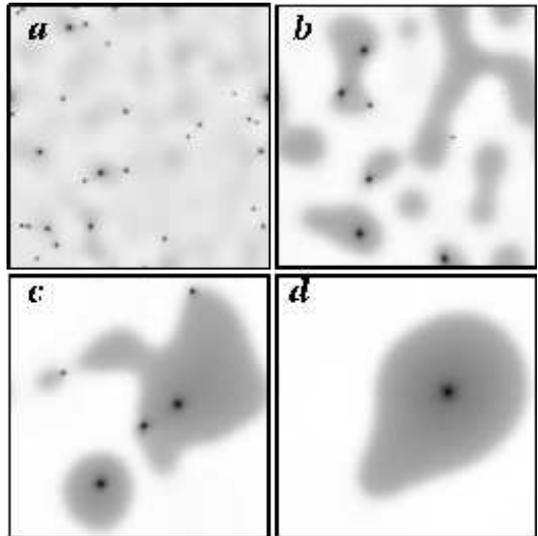,height=2.8in}}
\caption{Evolution of the field $|\psi|$, vortices are shown as black dots, 
white area corresponds to larger $|\psi|$ (and $\rho$), dark vise versa. 
Parameters: $\bar \xi=1, \xi_2 = 1$, $\beta =0.04$,$f_0=1.5-|\psi|^2, 
f_1= 1.4-0.7 \rho$, $
\delta=0.3$,$  
\rho_0=0.25, \alpha=0.03$. Images are shown for $t=80 (a), 400 (b), 
1600 (c) $ and $t=3200 (d)$. } 
\label{Fig1}
\end{figure}

\begin{figure}[h]
\centerline{  \psfig{figure=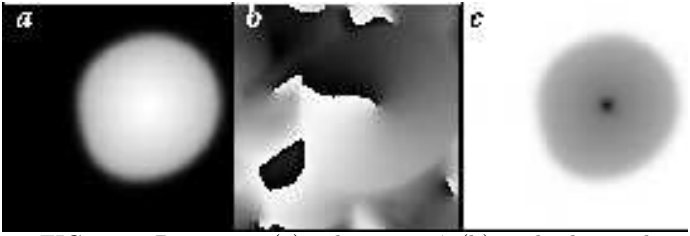,height=1.2in}}
\caption{ Density $\rho$ 
(a),  phase $\arg \psi$ (b) and tilt amplitude $|\psi|$ (c)  
for $t=4960$, other parameters as in  Fig. \protect \ref{Fig1}. Black
corresponds to $\rho=0=\arg \psi=|\psi|=0$, white to $\rho=1.4,
\arg \psi=2\pi, |\psi|=1.2$. }
\label{Fig1_1}
\end{figure}

At the initial stage of the evolution many vortices and small dense
clusters (islands) are created throughout the domain of integration
(Fig. \ref{Fig1}a). Islands are seen as  darker areas on the figure
because an increase in density $\rho$ results in the decrease of the
amplitude of $\psi$.  Some islands trap vortices and are practically
immobile, others don't contain vortices and drift in the direction
defined by the mean value of the tilt ${\bf n}$.  With time,  small
islands disappear and bigger islands grow (Fig. \ref{Fig1}b,c). 
It is interesting to note that due to tilt-driven drift coarsening
occurs much faster than in ordinary Cahn-Hilliard dynamics.
Finally, one big
island with the vortex in the center is formed (Fig. \ref{Fig1}d).
Corresponding density $\rho$ and the phase field $\arg \psi$ are shown
in Fig.  \ref{Fig1_1}. Surprisingly, even far from the island there some
``dormant vortices'' in the low-density phase (gas). These vortices are
seen as the end-points of the phase singularity lines (lines between
dark and white)  in Fig. \ref{Fig1_1}b.  These vortices do not
annihilate because in the low-density phase the diffusion terms in Eq.
(\ref{gle}) are absent.  They are not seen  in  Fig. \ref{Fig1_1}c because
their core size is small in the gas phase.  However, these vortices
can be advected into the
high-density phase from the edges. 

\begin{figure}[h]
\centerline{ \psfig{figure=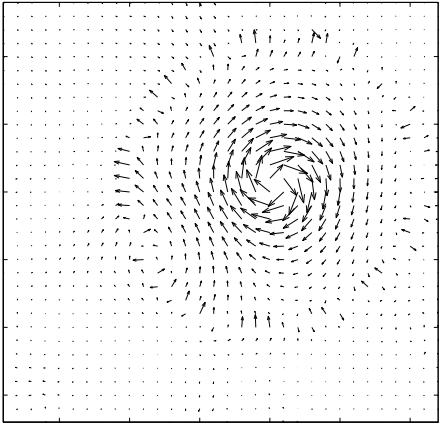,height=2.4in}}
\caption{ Velocity field corresponding to Fig. \protect \ref{Fig1_1}.}
\label{Fig2}
\end{figure}

\begin{figure}[h]
\centerline{ \psfig{figure=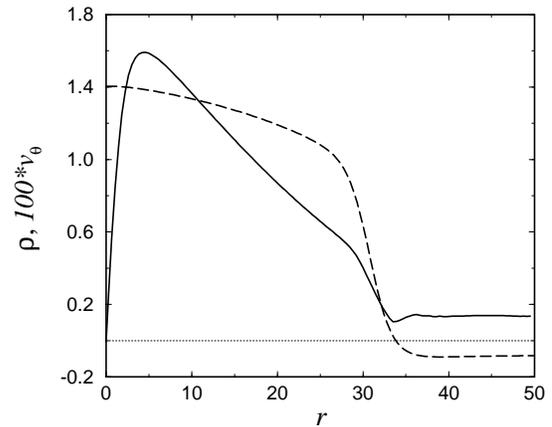,height=2.5 in}}
\caption{ Azimuthal velocity $v_\theta$ (solid line)  and 
density $\rho$ (dashed line) vs $r$ for the parameter of 
Fig. \protect \ref{Fig1}.} 
\label{Fig3}
\end{figure}

The velocity field is calculated using Eq. (\ref{vel}).
Fig. \ref{Fig2} shows the velocity field corresponding to the calculated
tilt structure of Fig.\ref{Fig1_1}. The rods perform circular motion around the 
vortex center. 
The azimuthal  velocity $v_\theta$ vs $r$ is shown in Fig. \ref{Fig3}. 
The velocity is maximal somewhere between the core and the island edge. 
It qualitatively  resembles  the experimental one \cite{kudrolli}. 
Outside the island the tilt becomes large and  the velocity
becomes small. 

In conclusion, we developed  a phenomenological model of the
formation  of the vortical ordered state in the
system of vertically vibrated rods. Our continuum model is based on a 
Cahn-Hilliard equation for the rods aerial density coupled to the
Ginzburg-Landau equation for the rod tilt. The model 
reproduces the qualitative features of the vortex formation process 
observed in the recent experiment  \cite{kudrolli}. We thank Toni 
Neicu and Arshad Kudrolli for stimulating discussions. 
This work was supported by the U.S.Department of Energy under grants
W-31-109-ENG-38 and DE-FG03-95ER14516.
Simulations were performed at the National Energy Research
Scientific Computing Center.

\end{document}